\documentclass[12pt]{iopart}
\usepackage{amsfonts}
\usepackage{amssymb}
\usepackage{graphicx}

\begin{document}

\title[Long rods on a square lattice]{Entropy-driven phase transition in a system of long rods on a square lattice}

\author{D H Linares$^1$, F Rom\'a$^{1,2}$ and A J Ramirez-Pastor$^1$}

\address{1 Departamento de F\'{\i}sica, Instituto de F\'{\i}sica Aplicada, Universidad Nacional de San
Luis-CONICET, Chacabuco 917, 5700 San Luis, Argentina}
\address{2 Centro At\'omico Bariloche, Av. Bustillo 9500,
8400 S. C. de Bariloche, Argentina}

\ead{antorami@unsl.edu.ar}

\begin{abstract}
The isotropic-nematic (I-N) phase transition in a system of long
straight rigid rods of length $k$ on square lattices is studied by
combining Monte Carlo simulations and theoretical analysis. The
process is analyzed by comparing the configurational entropy of
the system with the corresponding to a fully aligned system, whose
calculation reduces to the 1D case. The results obtained $(1)$
allow to estimate the minimum value of $k$ which leads to the
formation of a nematic phase and provide an interesting
interpretation of this critical value; $(2)$ provide numerical
evidence on the existence of a second phase transition (from a
nematic to a non-nematic state) occurring at density close to $1$
and $(3)$ allow to test the predictions of the main theoretical
models developed to treat the polymers adsorption problem.

\end{abstract}

\pacs{05.50.+q, 64.70.Md, 75.40.Mg} \submitto{JOURNAL OF
STATISTICAL MECHANICS: THEORY AND EXPERIMENTS}
\maketitle

\section{Introduction}

The study of systems of hard non-spherical colloidal particles has
been an attractive and important topic in statistical physics for
a long time. In the years 1941-1961, several papers contributed
greatly to our understanding of this
field~\cite{FLORY,HUGGI,MILLER,GUGGE,ONSAGER,ZIMM,ISIHARA,DIMA}.
Among them, Onsager~\cite{ONSAGER} predicted that very long and
thin rods interacting with only excluded volume interaction can
lead to long-range orientational (nematic) order. This nematic
phase, characterized by a big domain of parallel molecules, is
separated from an isotropic state by a phase transition occurring
at a finite critical density. Flory~\cite{FLORY} studied a lattice
model of long rod-like molecules, and based on a mean-field
approximation, argued that the lattice model would also show an
isotropic-nematic (I-N) phase transition as a function of density.
Later, DiMarzio~\cite{DIMA} gave an answer to the question of how
many ways we can pack together $N$ linear polymers in $M$ sites,
given a definite distribution of shapes for the molecules and a
definite distribution (continuous or discrete) of orientations of
each shape. The detailed knowledge of the orientations of the
molecules allowed to identify the various types (nematic, smetic,
and cholestic) of liquid crystals.
Refs.~\cite{FLORY,HUGGI,GUGGE,DIMA} will be discussed in more
detail in Section 4.

The phase properties of systems with purely steric interactions
are important from a statistical mechanical perspective because
the potential energy, $U$, of a steric system is, by definition,
constant. Consequently, the Helmholtz free energy $F = U -TS$ is
controlled by entropy ($S$) alone and all phase transitions are
entropy driven. The problem proposed by Onsager is a clear example
of an entropy-driven phase transition. Other examples,
corresponding to phase transitions in hard sphere systems, can be
found in Refs.~\cite{BRIDG,RICE}.

Despite of the physical relevance of such issues, rigorous results
are still very limited. An interesting overview of this topic can
be found in the work by Ioffe et al.~\cite{IOFFE} and references
therein. In this paper, the authors studied a system of rods on
$\mathbb{Z}^2$ with hard-core exclusion, each rod having a length
between $2$ and $N$. The existence of a I-N phase transition was
rigorously proved for sufficiently large $N$ and suitable
fugacity.

On the other hand, numerous experimental and numerical studies
have been recently devoted to the analysis of phase transitions in
systems of non-spherical
particles~\cite{VIAMONTES,DEMICHELE,VINK,CUETOS,GHOSH,MLR}. Of
special interest are those studies dealing with lattice versions
of this problem, where the situation is much less clear. In this
sense, a system of straight rigid rods of length $k$ on a square
lattice, with two allowed orientations, was studied in the
excellent paper by Ghosh and Dhar~\cite{GHOSH}.  The authors found
strong numerical evidence that the system shows nematic order at
intermediate densities for $k \geq 7$ and provided a qualitative
description of a second phase transition (from a nematic order to
a non-nematic state) occurring at a density close to $1$ (figure 1
in Ref.~\cite{GHOSH} shows a schematic representation of the
different phases corresponding to a system of long rigid rods on a
square lattice). As it was recently confirmed~\cite{MLR}, the
low-density I-N phase transition of rigid rods on square lattices,
with two allowed orientations, belongs to the 2D Ising
universality class.

Similar results have been obtained in previous
work~\cite{IOFFE,GAUTAM1,GAUTAM2,GAUTAM3}, which indicate the
universal behavior of these models. In the case of the system
discussed in Ref.~\cite{IOFFE} (see third paragraph), the authors
studied the problem of infinite $N$ (which can be mapped onto a 2D
Ising model), and showed that the case of finite (but large) $N$
can be seen as a perturbation of the latter. It is then reasonable
to expect that the finite $N$ case belongs to the same
universality class as the $N = \infty$. An Ising-like phase
transition was also found in a lattice system of semiflexible
living polymers~\cite{GAUTAM1,GAUTAM2,GAUTAM3}.


Even though the universality class of a system of straight rigid
rods of length $k$ on a square lattice (with two allowed
orientations) has been resolved~\cite{MLR}, other aspects of the
problem remain still poorly understood. Among them, the minimum
value of $k$($k_{min}=7$), which allows the formation of a nematic
phase, has been estimated from the behavior of a geometric order
parameter, without any theoretical justification and the second
phase transition predicted by Ghosh and Dhar~\cite{GHOSH} has not
been rigorously proved yet. In this context, the objectives of the
present work are to shed light on the underlying physics of the
observed $k$ dependence of the I-N phase transition, and to
contribute to the discussion on the existence or non-existence of
a second phase transition in the system. For this purpose, the
configurational entropy of a system of rigid rods deposited on a
square lattice is calculated by Monte Carlo (MC) simulations and
thermodynamic integration method ~\cite{BINDER1}. The numerical
data are compared with the corresponding ones obtained from a
fully aligned system (nematic phase), whose calculation reduces to
the one-dimensional case~\cite{PRB3}. The study allows to
calculate $k_{min}$, corroborating the previous result in
Ref.~\cite{GHOSH}, and provides an interesting interpretation of
this value of $k$. The results obtained provide also numerical
evidence on the existence of a second phase transition occurring
at high density.

Finally, the transition is studied from the main theoretical
models developed to treat the polymers adsorption
problem~\cite{LANG9,FLORY,HUGGI,GUGGE,DIMA,LANG11,IJMP}. Three
theories have been compared with the Monte Carlo data: the first
is the well-known Flory-Huggins (FH)
approximation~\cite{LANG9,FLORY,HUGGI}; the second is the
Guggenheim-DiMarzio (GD) approach for rigid rod
molecules~\cite{GUGGE,DIMA}; and the third is the recently
developed Semiempirical Model for Adsorption of Polyatomics (SE),
which is a combination of exact 1D calculations and GD
approximation~\cite{LANG11,IJMP}. The comparison indicates that
the SE model leads to an approximation significantly better than
the other existing approaches.

\section{Model and Monte Carlo method}

We address the general case of adsorbates assumed to be linear
rigid particles containing $k$ identical units ($k$-mers), with
each one occupying a lattice site. Small adsorbates with spherical
symmetry would correspond to the monomer limit ($k = 1$). The
distance between $k$-mer units is assumed to be equal to the
lattice constant; hence exactly $k$ sites are occupied by a
$k$-mer when adsorbed. The only interaction between different rods
is hard-core exclusion: no site can be occupied by more than one
$k$-mer. The surface is represented as an array of $M = L \times
L$ adsorptive sites in a square lattice arrangement, where $L$
denotes the linear size of the array.

Configurational entropy was calculated by using MC simulations and
thermodynamic integration
method~\cite{BINDER1,HANSEN,BINDER2,BINDER3,POLGREEN,LANG6}. The
method in the grand canonical ensemble relies upon integration of
the chemical potential $\mu$ on coverage along a reversible path
between an arbitrary reference state and the desired state of the
system. This calculation also requires the knowledge of the total
energy $U$ for each obtained coverage. Thus, for a system made of
$N$ particles on $M$ lattice sites, we have:

\begin{equation}
S(N,M,T)=S(N_0,M,T)+{U(N,M,T)-U(N_0,M,T) \over T} - {1 \over T}
\int_{N_0}^N{ \mu dN'}.\label{entN}
\end{equation}
In our case $U(N,M,T)=0$ and the determination of the entropy in
the reference state, $S(N_0,M,T)$, is trivial [$S(N_0,M,T)=0$ for
$N_0 = 0$]. Note that the reference state, $N \rightarrow 0$, is
obtained for $\mu/k_B T \rightarrow -\infty$. Then,
\begin{equation}
{s(\theta,T) \over k_B}= - {1 \over k_B T} \int_{0}^{\theta}{
\frac{\mu}{k} ~ d \theta'}\label{ent}
\end{equation}
where $s(=S/M)$ is the configurational entropy per site,
$\theta(=k~N/M)$ is the surface coverage (or density) and $k_B$ is
the Boltzmann constant.

In order to obtain the curve of $\mu$ vs $\theta$, a typical
adsorption-desorption algorithm in grand canonical ensemble has
been used~\cite{LANG11,LANG7,LANG10,SS12,PHYSA14}. The MC
procedure is as follows. Once the value of $\mu/k_BT$ is set, a
linear $k$-uple of nearest-neighbor sites is chosen at random.
Then, if the $k$ sites are empty, an attempt is made to deposit a
rod with probability $W={\rm min} \left\{1,\exp\left( \mu/k_BT
\right) \right\}$; if the $k$ sites are occupied by units
belonging to the same $k$-mer, an attempt is made to desorb this
$k$-mer with probability $W={\rm min} \left\{1,\exp\left(
-\mu/k_BT \right) \right\}$; and otherwise, the attempt is
rejected. In addition, displacement (diffusional relaxation) of
adparticles to nearest-neighbor positions, by either jumps along
the $k$-mer axis or reptation by rotation around the $k$-mer end,
must be allowed in order to reach equilibrium in a reasonable
time. A MC step (MCs) is achieved when $M$ $k$-uples of sites have
been tested to change its occupancy state. Typically, the
equilibrium state can be well reproduced after discarding the
first $r'=10^6$ MCs. Then, the next $r=2 \times 10^6$ MCs are used
to compute averages.


In our MC simulations, we varied the chemical potential and
monitored the density $\theta$, which can be calculated as a
simple average over the $r$ MC simulation runs. All calculations
were carried out using the parallel cluster BACO of Universidad
Nacional de San Luis, Argentina. This facility consists of 60 PCs
each with 3.0 GHz Pentium-4 processors.

\section{Numerical results}

Computational simulations have been developed for a system of
straight rigid rods of length $k$ ($k=2-10$) on a lattice. The
surface was represented as an array of adsorptive sites in a
square $L \times L$ arrangement with $L/k=20$, and periodic
boundary conditions. As we will show in figure 1, finite-size
effects are small for the coverage dependence of the chemical
potential with this lattice size.

\begin{figure}
\begin{center}
\includegraphics[width=9cm,clip=true]{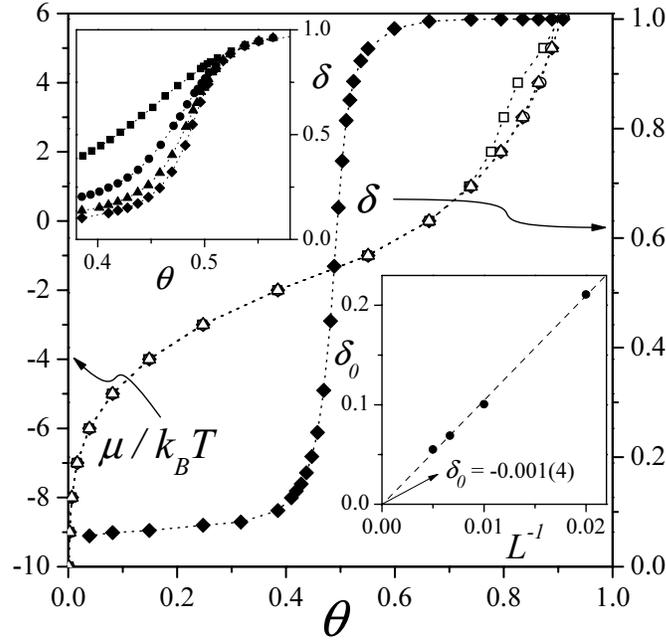}
\caption{Nematic order parameter (full diamonds, right axis) for
$k=10$ and $L/k=20$ and surface coverage dependence of the
chemical potential (left axis) for $k=10$ and  different lattice
sizes: $L/k=5$, open squares; $L/k=10$, open circles; $L/k=15$,
open triangles; and $L/k=20$, open diamonds. Upper-left inset:
Size dependence of the order parameter as a function of coverage.
Symbols are: squares, $L/k=5$; circles, $L/k=10$; triangles,
$L/k=15$; and diamonds, $L/k=20$. Lower-right inset: Order
parameter in the isotropic phase, $\delta_0$, as a function of
$L^{-1}$. The dashed line corresponds to a linear fit of the
data.} \label{f.1}
\end{center}
\end{figure}

\begin{figure}
\begin{center}
\includegraphics[width=9cm,clip=true]{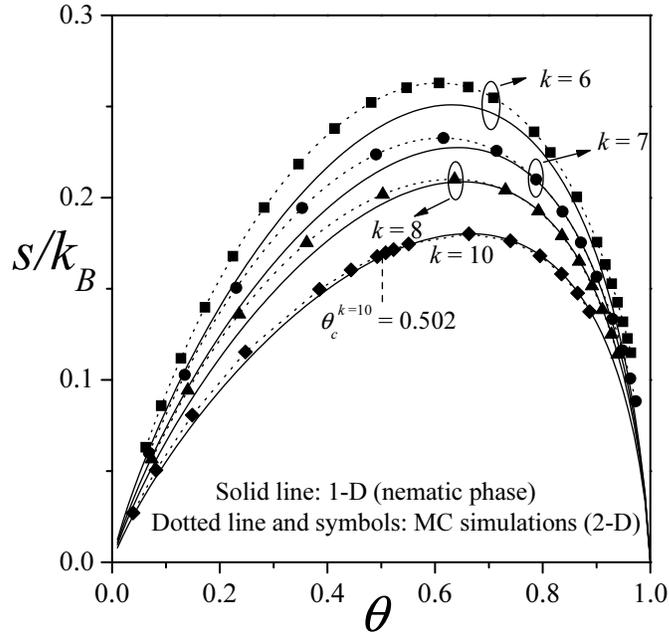}
\caption{ Configurational entropy per site (in units of $k_B$)
versus surface coverage for different $k$ as indicated. Dotted
line and symbols represent MC results for square lattices, and
solid lines correspond to exact results for one-dimensional
systems.} \label{f.2}
\end{center}
\end{figure}

The calculation of $s(\theta)/k_B$ through eq.~(\ref{ent}) is
straightforward and computationally simple, since the coverage
dependence of $\mu/k_BT$ is evaluated following the standard
procedure of MC simulation described in previous section. Then,
$\mu(\theta)/k_BT$ is spline-fitted and numerically integrated.
Typical curves of $\mu/k_BT$ vs $\theta$, obtained for $k=10$ and
different values of $L/k$ ($L/k=5$, open squares; $L/k=10$, open
circles; $L/k=15$, open triangles; and $L/k=20$, open diamonds),
are depicted in figure 1. In all cases, a smooth coverage
dependence is observed as result of the large number of averaged
configurations. The data collapse in a single curve for small and
intermediate values of the coverage ($\theta < 0.7$); however, the
disagreement turns out to be significantly large for larger
$\theta$'s. This difference diminishes when $L/k$ is increased,
being negligible for $L/k > 10$.

Figure 1 also shows the nematic order parameter
$\delta$~\cite{FOOT1} as a function of the coverage (full
diamonds). In this case, the data correspond to $k=10$ and
$L=200$. When the system is disordered ($\theta<\theta^c_1$,
$\theta^c_1$ being the critical coverage characterizing the I-N
phase transition at intermediate densities), all orientations are
equivalents and $\delta$ tends to zero. As the density is
increased above $\theta^c_1$, the $k$-mers align along one
direction and $\delta$ is different from zero. The behavior of
$\delta$ is a clear evidence of the existence of a I-N phase
transition in the system~\cite{GHOSH,MLR}. As a difference with
respect to the behavior of $\mu/k_BT$ vs $\theta$, important
finite-size effects are observed in $\delta(\theta)$ (see inset in
the upper-left corner of the figure, where $\delta$ is plotted for
different values of $L/k$: $L/k=5$, squares; $L/k=10$, circles;
$L/k=15$, triangles; and $L/k=20$, diamonds). This is the reason
for which the order parameter does not vanish at low coverage. In
order to clarify this point, a simple finite-size scaling analysis
is presented in the lower-right inset of the figure. In this
study, the limit value of the order parameter in the isotropic
phase, $\delta_0$, is plotted as a function of $L^{-1}$. This
limit value was obtained for $\theta \approx 0.1$. As it can be
observed, $\delta_0 \rightarrow 0$ as $ L \rightarrow \infty$.

Results of thermodynamic integration are shown in figure 2 for
straight rigid rods of different sizes ($k=6$, squares; $k=7$,
circles; $k=8$, triangles and $k=10$, diamonds) on square
lattices. Other sizes are not shown for clarity. The general
features of the coverage dependence of the entropy per site are
the following: in the limit $\theta=0$ the entropy tends to zero.
For very low densities, $s(\theta)/k_B$ is an increasing function
of $\theta$, reaches a maximum at $\theta_m$, then decreases
monotonically to a finite value for $\theta=1$. The position of
$\theta_m$ shifts to higher coverage as  $k$ gets larger. The
overall effect of the adsorbate size is to decrease the entropy
for all coverage.


For $\theta^c_1< \theta<\theta^c_2$, $\theta^c_2$ being the
critical coverage characterizing the second phase transition (from
a nematic order to a non-nematic state) occurring at a density
close to $1$, the system is characterized by a big domain of
parallel $k$-mers. The calculation of the entropy of this fully
aligned state having density $\theta$ reduces to the calculation
of a one-dimensional problem~\cite{PRB3}
\begin{equation}
{s(\theta) \over k_B} =  \left[1-{\left(k-1 \right) \over k}
\theta \right]\ln \left[1-{\left(k-1 \right) \over k} \theta
\right]- {\theta \over k}\ln{\theta \over k} -\left(1-\theta
\right)\ln \left(1-\theta \right). \label{s1d}
\end{equation}

Results from eq.~(\ref{s1d}) for different adsorbate sizes are
shown in figure 2 (solid lines). Interesting conclusions can be
drawn from the figure. Namely, for $k \leq 6$, the 1D results
present a smaller $s/k_B$ than the 2D simulation data over all the
range of $\theta$. For $k \geq 7$, there exists a range of
coverage for which the difference between the 1D value and the
true 2D value is very small. In other words, for $k \geq 7$ and
intermediate densities, it is more favorable for the rods to align
spontaneously, since the resulting loss of orientational entropy
is by far compensated by the gain of translational entropy. These
results corroborate the previous results in Ref.~\cite{GHOSH}, and
provides a physical interpretation of this critical value of $k$.
In addition, the technique supply an alternative method of
determining the critical coverage characterizing the I-N phase
transition. In fact, $\theta^c_1$ can be calculated from the
minimum value of $\theta$ for which occurs the near superposition
of the 1D and 2D results. As an example, the curves for $k=10$
lead to $\theta^c_1 \approx 0.5$ in good agreement with the
recently reported value of $\theta^c_1 \approx
0.502(1)$~\cite{MLR}. Thus, the simulation scheme presented here
appears as a simple method to approximate $\theta^c_1$ without any
special requirement and time consuming computation. However, it is
important to emphasize that the calculation of the entropy of the
nematic phase from the 1D model is an approximation (especially at
the moderate densities, where the phase is not completely
aligned). Consequently, a precise determination of $\theta^c_1$
should require an extensive work of MC simulation and finite-size
scaling techniques~\cite{MLR}.

On the other hand, the results in figure 2 provide valuable
information about the second phase transition predicted in
Ref.~\cite{GHOSH}. In fact, figure 3 shows the behavior of one of
the curves in figure 2 (that corresponding to $k=7$) at density
close to $1$. As it can be observed, the configurational entropy
of the 2D system (solid circles) differs from the corresponding
one to the nematic phase (solid line). The exact density where the
two curves separate should be indicative of $\theta^c_2$. Even
though the determination of this point is out of the scope of the
present paper, it is clear that in this case $\theta^c_2$ varies
between $0.87$ and $0.93$ (as indicated with arrows).

As it was mentioned in Ref.~\cite{GHOSH}, the relaxation time
increases very fast as the density increases. Consequently, MC
simulations at high density are very time consuming and may
produce artifacts related to non-accurate equilibrium states. In
order to discard this possibility, we study the effect of the
number of MCs used in the calculations on the behavior of $s/k_B$
at high density. For this purpose, the curve obtained for
equilibration times of the order $O(10^7 MCs)$ (solid circles),
with an effort reaching almost the limits of our computational
capabilities, is compared with the corresponding one obtained for
$r'=r=10^4$ MCs (open circles). The tendency is very clear: the
better the statistics, the larger the separation between line and
symbols. Other intermediate cases, not shown for sake of clarity,
reinforce the last argument. Then, the difference between the 1D
and 2D results can not be associated to numerical
limitations~\cite{FOOT2} and is a clear evidence of the existence
of a second phase transition at high coverage.

Finally, the measurement of the nematic order parameter as a
function of the coverage was used as an independent method to
characterize the phase transition occurring in the system (see
figure 4). The calculations were developed for $k=7$, $L=140$ and
$r'=r = 10^7$ MCs. The behavior of $\delta$ is also indicative of
the existence of a transition from a nematic to a non-nematic
state. As in figure 1, a simple finite-size scaling analysis of
the limit value of the order parameter in the non-nematic phase
shows that $\delta_0 \rightarrow 0$ as $ L \rightarrow \infty$
(see inset). In this case, the limit value was obtained for
$\theta \approx 0.96$. The robustness and consistency of the
analysis presented in figures 3 and 4 allow us to confirm the
existence of a second phase transition (from a nematic order to a
non-nematic state) occurring in a system of rigid rods on square
lattices at high density. In addition, this analysis provides the
first numerical evidence existing in the literature about this
important point.

\begin{figure}
\begin{center}
\includegraphics[width=9cm,clip=true]{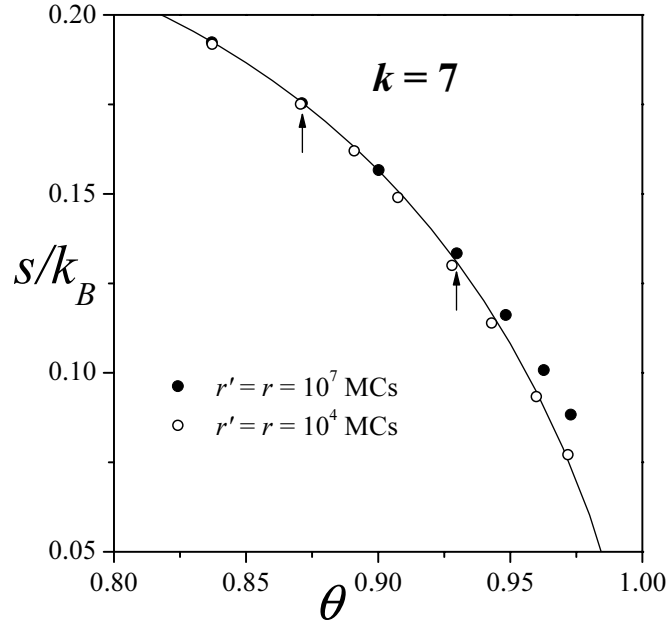}
\caption{Configurational entropy per site (in units of $k_B$)
versus surface coverage for $k=7$ and high densities. Curves
correspond to 1D system (solid line), 2D system with $r'=r = 10^7$
MCs (solid circles) and 2D system with $r'=r = 10^4$ MCs (open
circles).} \label{f.3}
\end{center}
\end{figure}

\begin{figure}
\begin{center}
\includegraphics[width=9cm,clip=true]{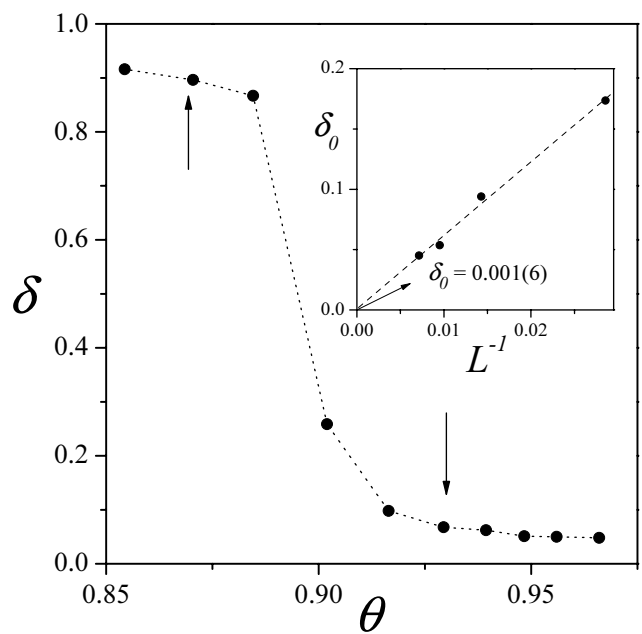}
\caption{Nematic order parameter as a function of coverage for
$k=7$ and $r'=r = 10^7$ MCs. Inset: Order parameter in the
isotropic phase, $\delta_0$, as a function of $L^{-1}$. The dashed
line corresponds to a linear fit of the data.  } \label{f.4}
\end{center}
\end{figure}

\begin{figure}
\begin{center}
\includegraphics[width=9cm,clip=true]{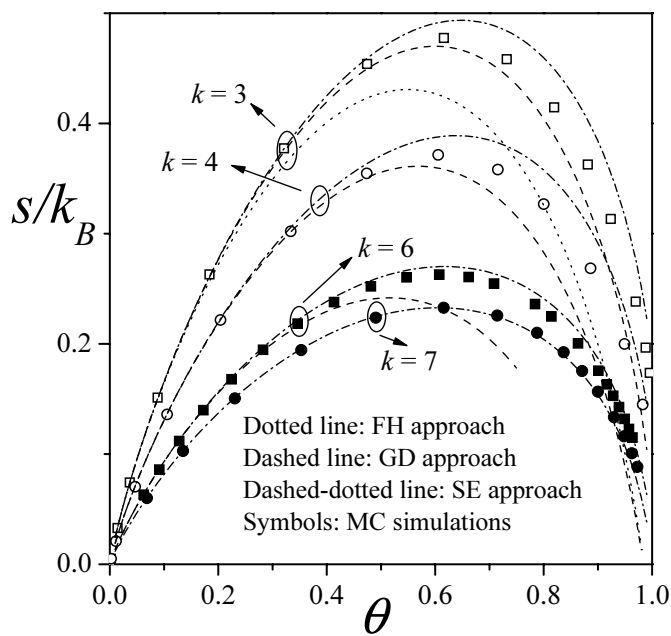}
\caption{Configurational entropy per site (in units of $k_B$)
versus surface coverage for different values of $k$ as indicated.
The symbols are indicated in the figure.} \label{f.5}
\end{center}
\end{figure}

\section{Analytical approximations and comparison between simulated and theoretical results}

Next, the transition is studied from the main theoretical models
developed to treat the polymers adsorption
problem~\cite{FLORY,HUGGI,GUGGE,DIMA,LANG9,LANG11,IJMP}. The study
allows us to investigate the predictions, reaches and limitations
of those theories when are used to describe the I-N phase
transition occurring in a system of long rods on a lattice.

Three theories have been considered: the first is the well-known
Flory-Huggins (FH) approximation~\cite{FLORY,HUGGI}; the second is
the Guggenheim-DiMarzio (GD) approach for rigid rod
molecules~\cite{GUGGE,DIMA}; and the third is the recently
developed Semiempirical Model for Adsorption of Polyatomics
(SE)~\cite{LANG11,IJMP}.

As it was mentioned in Section 1, Onsager~\cite{ONSAGER},
Zimm~\cite{ZIMM} and Isihara~\cite{ISIHARA} made important
contributions to the understanding of the statistics of rigid rods
in dilute solution. These treatments are limited in their
application because they are valid for dilute solution only and
because they are not applicable to systems of non-simple shapes.
The FH theory, due independently to Flory~\cite{FLORY} and to
Huggins~\cite{HUGGI}, has overcome the restriction to dilute
solution by means of a lattice calculation. The approach is a
direct generalization of the theory of binary liquids in two
dimensions or polymer molecules diluted in a monomeric solvent. It
is worth mentioning that, in the framework of the lattice-gas
approach, the adsorption of $k$-mers on homogeneous surfaces is an
isomorphous problem to the binary solutions of polymer-monomeric
solvent.

A great deal of work has been done on checking the predictions of
the FH theory against experimental results, being the theory
completely satisfactory in a qualitative, or semi-quantitative
way. There is no doubt that this simple theory contains the
essential features which distinguish high polymer solutions from
ordinary solutions os small molecules. In the framework of the FH
approach, the configurational entropy per site can be written
as~\cite{IJMP}:
\begin{equation}
{s(\theta) \over k_B} =  - {\theta \over k}\ln{\theta \over k}
-\left(1-\theta \right)\ln \left(1-\theta \right) - {\theta \over
k} \left[k-1- \ln \left( c \over 2 \right) \right] \  \  \  \  \ \
\ {\rm (FH) \label{fh}}.
\end{equation}

The FH statistics, given for the packing of molecules of arbitrary
shape but isotropic distribution, provides a natural foundation
onto which the effect of the orientation of the ad-molecules can
be added. Following this line of thought, DiMarzio~\cite{DIMA}
developed an approximate method of counting the number of ways,
$\Omega$, to pack together linear polymer molecules of arbitrary
shape and of arbitrary orientations. Accordingly, $\Omega$ was
evaluated as a function of the number of molecules in each
permitted direction. These permitted directions can be continuous
so that $\Omega$ is derived as a function of the continuous
function $f(r)$ which gives the density of rods lying in the solid
angle $\Delta r$, or the permitted directions can be discrete so
that $\Omega$ is the the number of ways to pack molecules onto a
lattice. Based on the detailed knowledge of the orientations of
the molecules, the various types (nematic, smetic, and cholestic)
of liquid crystals were argued for and the reasons for their
existence were ascertained. In the case of allowing only those
orientations for which the molecules fit exactly onto the lattice
is that for the case of an isotropic distribution the value of
$\Omega$ reduces to the value obtained previously by
Guggenheim~\cite{GUGGE}. In this limit, which we call
Guggenheim-DiMarzio approximation, the corresponding expression
for configurational entropy per site is:
\begin{eqnarray}
{s(\theta) \over k_B} & = &  - {\theta \over k}\ln{\theta \over k}
-\left(1-\theta \right)\ln \left(1-\theta \right) + \left(\theta-
{c \over 2} \right) \ln \left( c \over 2 \right)      \nonumber\\
& & + \left[{c \over 2}-{\left(k-1 \right) \over k} \theta
\right]\ln \left[{c \over 2}-{\left(k-1 \right) \over k} \theta
\right] \ \ \ \ \ \ \ \ \ \ \ \ \ \ \ \ \ {\rm (GD) \label{gd}}.
\end{eqnarray}

More recently, a new theory to describe adsorption of rigid rods
has been introduced. The model, hereafter denoted EA, is based on
exact forms of the thermodynamic functions of linear adsorbates in
one dimension and its generalization to higher
dimensions~\cite{IJMP}. Detailed comparisons between theoretical
and simulation results of adsorption~\cite{LANG11} shown that GD
approach fits very well the numerical data at low coverage, while
EA model behaves excellently at high coverage. Based on these
findings, the Semiempirical Model for Adsorption of Polyatomics
was developed~\cite{LANG11,IJMP}. SE model is a combination of
exact calculations in 1D and GD approximation with adequate
weights. In this approach, the configurational entropy per site is
given by:
\begin{eqnarray}
{s(\theta) \over k_B} & = &  - {\theta \over k}\ln{\theta \over k}
-\left(1-\theta \right)\ln \left(1-\theta \right) + \theta
\left[{1 \over 2} - {c \over 4} + {1 \over k} \ln \left( c \over 2
\right) \right] \nonumber\\
& & +{1 \over 2} {k \over \left(k-1 \right)} \left[1-{\left(k-1
\right)^2 \over k^2} \theta^2 \right]\ln \left[1-{\left(k-1
\right) \over k} \theta \right] \nonumber\\
& & - {c \over 4} \left[\theta + {k \left(c-4 \right)+4 \over 2
\left(k-1 \right)} \right] \left[1-{2 \left(k-1 \right) \over c k}
\theta \right]\ln \left[1-{2 \left(k-1 \right) \over c k} \theta
\right] \ \ \ \ \ \ \  {\rm (SE) \label{se}}.
\end{eqnarray}

The comparison between simulation data and theoretical predictions
for $3$-mers, $4$-mers, $6$-mers and $7$-mers adsorbed on square
lattices is shown in figure 5. The behavior of the analytical
approaches can be explained as follows. In all cases, the
agreement between simulation and theoretical data is very good for
small values of coverage. However, as the surface coverage is
increased, FH (dotted line) and GD (dashed line) predict a smaller
$s/k_B$ than the simulation data over the entire range of
coverage. With respect to the $k$-mer size, FH and GD become less
accurate as $k$ increases, being the disagreement significantly
large for $k \geq 3$ in the case of FH and $k \geq 6$ in the case
of GD. The behavior of SE (dashed-dotted line) is more complex.
For $k \leq 6$, the approach overestimate the value of the entropy
in the whole range of $\theta$; for $k=7$ the agreement between
simulation and theoretical data is excellent; and finally, for $k
\geq 8$ (data do not shown here for simplicity) the tendency is
inverted and the theoretical data present a smaller $s/k_B$ than
the simulation results. In summary, appreciable differences can be
seen for the different approximations studied in this
contribution, with SE being the most accurate for all sizes
considered here.

\begin{figure}
\begin{center}
\includegraphics[width=9cm,clip=true]{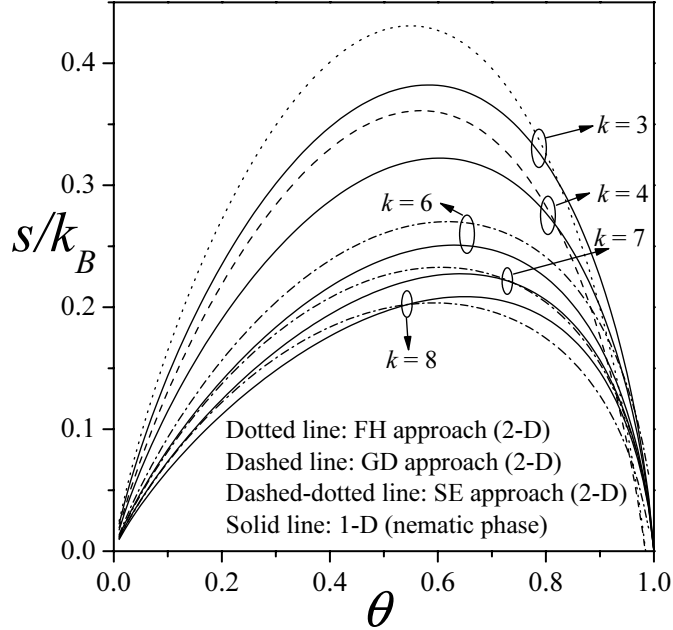}
\caption{Same as figure 2 for theoretical data. The symbols are
indicated in the figure.} \label{f.6}
\end{center}
\end{figure}

Once the theoretical approaches have been tested against numerical
experiments, comparisons between results from
eqs.~(\ref{fh}-\ref{se}) and the corresponding ones from
eq.~(\ref{s1d}) are shown in figure 6. In general, the behavior of
the 2D curves is the following. For low values of $k$, the 2D
system has higher entropy over all the range of coverage. From a
given value of $k$ (which depends on the approximation
considered), the 2D and 1D curves cross at intermediate densities
and two well differentiated regimes can be observed. In the first
regime, which occurs at low densities, the 2D approaches predict a
larger entropy than the 1D data. In the second regime (at high
densities) the behavior is inverted and the 2D data present a
smaller $s/k_B$ than the 1D results. Given that the theoretical
results in 2D assume isotropy in the adlayer (interested readers
are referred to Ref.~\cite{DIMA}, where this point is explicitly
considered), the crossing of the curves shows that, in the second
regime, the contribution to the 2D entropy from the isotropic
configurations is lower than the contribution from the aligned
states. Then, the existence of an intersection point is indicative
of a I-N transition and allows us to estimate $k_{min}$ and
$\theta^c_1$ from the different approximations studied.

As it can be observed, FH and GD approaches predict values of
$k_{min}=3$ and $k_{min}=4$, respectively. On the other hand, SE
approximation performs significantly better than the other
approaches, predicting the ``exact" value of
$k_{min}=7$~\cite{GHOSH}. With respect to $\theta^c_1$, although a
systematic analysis of the dependence of $\theta^c_1$ with $k$ was
not carried out since this was out of the scope of the present
work, it is interesting to note that, as it is
expected~\cite{GHOSH}, the three approximations predict that the
critical density decreases for increasing $k$.

In summary, we have addressed the critical properties of a system
of long straight rigid rods of length $k$ on square lattices. The
results were obtained by combining Monte Carlo simulations,
thermodynamic integration and theoretical analysis.

Two main conclusions can be drawn from the present work. On one
hand, the comparison between the configurational entropy of the
system and the corresponding to a fully aligned system confirms
previous results in the literature~\cite{GHOSH}, namely, the
presence of a I-N phase transition at intermediate densities for
$k \geq 7$.

On the other hand, the Monte Carlo study presented here represents
the first simulation evidence about the existence of a second
phase transition from a nematic order to a non-nematic state
occurring at high density. The determination of the order of this
second phase transition is still an open problem. An exhaustive
study on this subject, based on Monte Carlo simulations and
finite-size scaling theory, will be the object of future work.

\ack{This work has been supported by Universidad Nacional de San
Luis (Argentina) under project 322000, CONICET (Argentina) under
project PIP 6294 and the National Agency of Scientific and
Technological Promotion (Argentina) under project 33328 PICT
2005.}


\begin{thebibliography}{10}
\bibitem{FOOT1}
{$\delta$ is defined as in Refs.~\cite{GHOSH,MLR}: $\delta = \left
| v_1 -v_2 \right |/ \left ( v_1+v_2 \right )$, $v_1$ ($v_2$)
being the number of $k$-mers aligned along the horizontal
(vertical) direction}
\bibitem{FOOT2}
{That the 1D and 2D results differ for large densities has been
shown rigorously. Thus, while the 2D entropy has a non-zero lower
bound as the surface coverage tends to $1$, the 1D entropy tends
to zero (see, for instance, Refs.~\cite{IJMP} and~\cite{Wu}).}
\end{thebibliography}
\end{document}